\documentclass[twocolumn]{autart}

\usepackage{amsmath,amsfonts,amssymb}

\newcommand{\cal}{\mathcal}

\newfont{\gothic}{eufm10 scaled\magstep0}

\newcommand{\rr}{\mbox{$\mathbb R$}}
\newcommand{\nn}{\mbox{$\mathbb N$}}
\newcommand{\Sn}{\mbox{$\mathbb S$}}

\newtheorem{theorem}{Theorem}[section]
\newtheorem{lemma}[theorem]{Lemma}

\newtheorem{proposition}[theorem]{Proposition}

\newtheorem{definition}[theorem]{Definition}
\newtheorem{make_remark}[theorem]{Remark}

\begin{document}
\begin{frontmatter}

\title{ Correction to:\\
 ``Position estimation from direction or range measurements"
} 


\author[I3S,IUF]{Tarek Hamel}\ead{thamel@i3s.unice.fr},    
\author[I3S]{Minh-Duc Hua}\ead{huaminhduc@yahoo.com},
\author[INRIA,I3S]{Claude Samson}\ead{claude.samson@inria.fr, csamson@i3s.unice.fr}               

\address[I3S]{Universit\'e C\^ote d'Azur, CNRS, I3S, France}  
\address[INRIA]{Universit\'e C\^ote d'Azur, INRIA, France}
\address[IUF]{Institut Universitaire de France}

\begin{abstract}
This technical communiqu\'e aims at correcting an erroneous statement (Lemma 2.4) in an earlier paper \cite{HamSam2017} by the same authors concerning a sufficient condition of uniform observability for a Linear Time-Varying (LTV) system. In this earlier paper, the proofs of two other lemmas, about body-pose estimation from range measurements, relied on this erroneous statement. For the sake of conciseness, only a new proof of one of these lemmas is presented, the proof of the second lemma being a simpler version of it.
\end{abstract}

\end{frontmatter}

\section{Introduction}
The paper is organized as follows. Recalls of uniform observability for a LTV system, followed by propositions stating a necessary condition of non-uniform observability and two sufficient conditions of uniform observability, are presented in Section \ref{recalls}. At the end of this section, an example illustrates why the Lemma 2.4 in \cite{HamSam2017} is not correct in its actual form. Proofs given in \cite{HamSam2017} of two other lemmas, namely Lemmas 4.1 and 4.2, which state persistent excitation conditions that ensure uniform observability for body-pose estimation from range measurements, relied on this erroneous lemma. For this reason a new proof of the most advanced version of these lemmas, i.e. Lemma 4.2, is presented in Section \ref{proof_lemma4.2}. The simpler proof of Lemma 4.1 is easily obtainable by following the same lines and arguments.

\section{Conditions for uniform observability of a LTV system} \label{recalls}
Consider a generic linear time-varying (LTV) system
\vspace{-0.3cm}
\begin{equation} \label{linear_system}
\left\{ \begin{array}{lll}
\dot{X} & = & A(t)X+B(t)U \\
Y & = & C(t)X
\end{array} \right.
\vspace{-0.3cm}
\end{equation}
with $X \in \rr^n$ the system state vector, $U \in \rr^s$ the system input vector, and $Y \in \rr^m$ the system output vector.
\begin{definition}[uniform observability] \label{chen1984}
Sytem \eqref{linear_system} is {\em uniformly observable} if there exist $\delta>0$, $\mu>0$ such that $\forall t\geq 0$:
\vspace{-0.3cm}
{\small
\begin{equation} \label{grammian}
W(t,t+\delta):=\frac{1}{\delta} \int_t^{t+\delta}\Phi^{\top}(s,t)C^{\top}(s)C(s)\Phi(s,t)ds \geq \mu I_d >0
\vspace{-0.3cm}
\end{equation}
}
\end{definition}
with $\Phi(t,s)$ the transition matrix associated with $A(t)$, i.e. such that $\frac{d}{dt}\Phi(t,s)=A(t)\Phi(t,s)$ with $\Phi(t,t)=I_d$.
The matrix valued-function $W(t,t+\delta)$ is called the {\em observability Gramian} of System \eqref{linear_system}.

Let us further assume from now on that the k-th order time-derivative of the matrix-valued function $A$ (resp. $C$) is well defined and bounded on $[0,+\infty)$ up to $k=K\geq0$ (resp. up to $k=K+1$).\\
Define $N_0:=C$, $N_{k+1}:=N_{k}A+\dot{N}_k$, $k=1,\ldots$, and the set $\cal{M}_K$ of matrix-valued functions $M(.)$ of dimension $(q \times n)$ ($q \geq 1$) composed of row vectors of $N_0(.)$, $N_1(.)$,$\hdots$

\begin{proposition} \label{proposition2}
{\bf (necessary condition for non-uniform observability)}\\
Sytem \eqref{linear_system} is {\em not uniformly observable} only if the following statement
\begin{equation} \label{NUO}
\begin{array}{l}
\forall \delta>0, \exists \{t_p\}_{p\in \nn}, \exists x \in \Sn^{n-1}~:\\~\lim_{p \rightarrow +\infty}\int_0^{\delta} |M(t_p+s)\phi(t_p+s,t_p)x|^2 ds=0
\end{array}
\end{equation}
holds true for all matrix-valued functions in $\cal{M}_K$.
\end{proposition}
This proposition follows directly from \cite{Morin2017}, proof of Proposition 1, relation (16). It in turn yields the following proposition.

\begin{lemma} \label{lemma1}
The existence of a matrix $M \in \cal{M}_K$ satisfying the following property
\vspace{-0.5cm}
{\small
\begin{equation} \label{grammianbis}
\bar{W}(t,t+\delta):=\frac{1}{\delta} \int_t^{t+\delta}\Phi^{\top}(s,t)M^{\top}(s)M(s)\Phi(s,t)ds \geq \bar{\mu} I_d >0
\end{equation}
}
implies the satisfaction of \eqref{grammian}, and thus uniform observability of the corresponding LTV system.
\end{lemma}
{\bf Proof}\\
From Proposition \ref{proposition2}, if \eqref{grammian} were not true then there would exist a sequence $\{t_p \in \rr,~p\in \nn \}$ and a unit vector $x$ such that
\begin{equation} \label{lim}
\lim_{p \rightarrow +\infty}\int_{t_p}^{t_p+\delta} |M(s)\Phi(s,t_p)x|^2ds~=~0
\end{equation}
which would contradict \eqref{grammianbis}.

\begin{proposition} \label{proposition1}~\\
{\bf (sufficient conditions for uniform observability)}
Relation \eqref{grammianbis} is satisfied if any of the following properties holds true.

{\bf C1} (from \cite{Morin2017}, Proposition 1): There exists $M$ such that
\begin{equation} \label{det}
\frac{1}{\delta} \int_t^{t+\delta}|det \big( M ^{\top}(s)M(s) \big) |ds \geq \bar{\bar{\mu}} >0
\end{equation}
{\bf C2}: $A$ is a constant matrix with real eigenvalues, and there exists $M$ such that
\begin{equation} \label{M}
\frac{1}{\bar{\delta}} \int_t^{t+\bar{\delta}}M^{\top}(s)M(s)ds \geq \bar{\bar{\mu}} I_d >0
\end{equation}
\end{proposition}
The fact that {\bf C2} implies \eqref{grammianbis} is a direct consequence of Lemma 2.7 in \cite{HamSam2017} with the matrix $H$ of this lemma taken equal to the identity matrix.


{\bf Important remark}: The condition {\bf C2} suggests that the existence of a matrix-valued function $M$ that satisfies the inequality \eqref{M} {\em generically} entails the property of uniform observability. However, there are specific cases for which the satisfaction of this condition does not imply uniform observability of the LTV system. For instance, consider the the following state and output matrix-valued functions
\[
A=\left[ \begin{array}{cc}
0 & 1\\-1 & 0 \end{array} \right],~C(t)=\left[ \begin{array}{cc} \sin(t)^2 & 0.5\sin(2t) \\ 0.5\sin(2t) & \cos(t)^2 \end{array} \right]
\]
Note that the poles of $A$ are pure imaginary, so that {\bf C2} does not apply to this case, and that $C(t)$ is the operator projecting on the 2D-plane orthogonal to the vector $y(t) \equiv (\cos(t),-\sin(t))^{\top}$. The transition matrix associated with $A$ is
\[
\Phi(s,t)=exp(A(s-t))=\left[ \begin{array}{cc}
\cos(s-t) & \sin(s-t)\\-\sin(s-t) &  \cos(s-t)\end{array} \right]
\]
Consider the unit vector $x \equiv (1,0)^{\top}$. One easily verifies that $\Phi(s,0)x=y(s)$, and thus that
$C(s)\Phi(s,0)x$ is the null vector. This in turn implies that $\forall \delta>0,~x^{\top}W(0,\delta)x=0$ and proves that the corresponding LTV system is not uniformly observable. However, one also verifies that, by choosing $M(t)\equiv C(t)$
\[
\int_{t}^{t+\delta} M^{\top}(s)M(s) ds = \int_{t}^{t+\delta} C(s) ds
\]
is a positive matrix for any $\delta>0$. This example shows that the positivity of the last integral is not sufficient to establish uniform observability of the corresponding LTV system, and thus that Lemma 2.4 in \cite{HamSam2017} is not valid without adding complementary conditions, like in {\bf C2}.\\
The former proof of Lemma 4.2 in \cite{HamSam2017}, which is related to the estimation of a body pose from range measurements, is based on the (incorrect) Lemma 2.5.
We give next a correct proof of this Lemma, and leave the interested reader the task of verifying that a correct (simpler) proof of Lemma 4.1 in \cite{HamSam2017} is obtained by applying similar calculations and arguments.

\section{Proof of Lemma 4.2 in \cite{HamSam2017}} \label{proof_lemma4.2}
This lemma, and Lemma 4.1, concern the problem of estimating the (assumed bounded) position $x_{pos} \in \rr^n$ ($n=2$ or $n=3$) of a body equipped with sensors that measure the distance between the body and $l$ source points whose coordinates $z_i \in \rr^n$ ($i\in \{1,\hdots,l\}$) in the considered inertial frame are known. The (assumed bounded) body velocity $u(t) \in \rr^n$ is measured in the inertial frame, and the acceleration $\dot{u}(t) \in \rr^n$ is also assumed bounded. The difference with Lemma 4.1, is that it is further assumed in Lemma 4.2 that the velocity measurements are biased by some initially unknown additive component denoted by $a \in \rr^n$.
The lemma involves the following system's state, output, and input vectors:
\[
\begin{array}{l}
X:=[x_{pos}^{\top},a^{\top},y_0,a^{\top}x_{pos},|a|^2]^{\top} \\
Y:= [y_0,(y_1-y_0-0.5|z_1|^2),\hdots,(y_l-y_0-0.5|z_l|^2)]^{\top} \\
U:=[u^{\top},0_{1\times n},-\sum_{i=1}^l \alpha_i(z_i^{\top}u),0,0]^{\top}
\end{array}
\]
with
\[
\begin{array}{l}
y_0:=0.5|x_{pos}|^2-\sum_{i=1}^l\alpha_iz_i^{\top}x_{pos} \\
y_i:=0.5|x_{pos}-z_i|^2~,~~i\in \{1,\hdots,l\}
\end{array}
\]
and $\alpha_i:=[\alpha_1,\hdots,\alpha_l]^{\top}$ denoting a $l$-dimensional vector of real numbers such that $\sum_{i=1}^l\alpha_i=1$. The corresponding matrix-valued functions are:
\vspace{-0.3cm}
{\small
\[
A(t) = \left[ \begin{array}{ccccc} 0_{n\times n} & I_{n\times n} & 0_{n\times 1} & 0_{n\times 1} & 0_{n\times 1}\\
0_{n\times n} & 0_{n\times n} & 0_{n\times 1} & 0_{n\times 1} & 0_{n\times 1} \\
u^{\top}(t) & -\sum_{i=1}^l\alpha_i z_i^{\top} & 0 & 1 & 0 \\
0_{1\times n} & u^{\top}(t) & 0 & 0 & 1 \\
0_{1\times n} & 0_{1\times n} & 0 & 0 & 0
\end{array} \right]
\]}
\vspace{-0.3cm}
\[
C=\left[ \begin{array}{ccccc}
0_{1 \times n} & 0_{1 \times n} & 1 & 0 & 0 \\
D(\alpha)Z^{\top} & 0_{l \times n} & 0_{l \times 1} & 0_{l \times 1} & 0_{l \times 1}
\end{array} \right]
\]
with $Z:=[z_1 \ldots z_l]$, $D(\alpha):=\xi \alpha^{\top}-I_{l \times l}$ a matrix of rank $l-1$, and $\xi:=[1,\ldots,1]^{\top}$ a $l$-dimensional vector.
The input matrix $B$ is the identity matrix in this case.

The lemma 4.2 states that, if $\dot{u}(t)$ and the vectors $z_i$ ($i=1,\ldots,l$) satisfy the {\em persistently exciting} (p.e.) condition
\vspace{-0.3cm}
\begin{equation} \label{pecondition}
\forall t\geq 0:ZD^{\top}(\alpha)D(\alpha)Z^{\top}+\frac{1}{\delta}\int_{t}^{t+\delta}\dot{u}(s)\dot{u}^{\top}(s) ds \geq \mu I_d
\end{equation}
for some $\delta>0$ and $\mu>0$, then \eqref{grammian} holds true, thus implying that the condition of uniform observability is satisfied and that the Riccati observer proposed in \cite{HamSam2017} for the estimation of the body position is (locally) uniformly exponentially stable.\\
To prove this lemma we introduce the following matrix valued-function
\[
M(t):=\left[ \begin{array}{c} N_0 \\ N_1(t)\\ N_2(t)\end{array} \right]
\vspace{-0.3cm}
\]
with $N_0=C$, $N_1(t)=CA(t)$, $N_2(t)=N_1(t)A(t)+\dot{N}_1(t)$. Using the expressions of $A(t)$ and $C$ one verifies that
\vspace{-0.5cm}
{\small
\begin{equation} \label{matrixM}
M(t)=\left[ \begin{array}{ccccc}
0_{1 \times n} & 0_{1 \times n} & 1 & 0 & 0 \\
D(\alpha)Z^{\top} &  0_{l \times n} & 0_{l \times 1} & 0_{l \times 1} & 0_{l \times 1} \\
u^{\top}(t) & -\sum_{i=1}^l\alpha_i z_i^{\top} & 0 & 1 & 0 \\
0_{l \times n} & D(\alpha)Z^{\top} & 0 & 0 & 0 \\
\dot{u}(t)^{\top} & 2 u^{\top}(t) & 0 & 0 & 1
\end{array} \right]
\end{equation}}
One verifies that the transition matrix associated with $A(t)$ is
\vspace{-0.5cm}
{\small
\vspace{-0.5cm}
\begin{equation} \label{matrixPhi}
\phi(t+s,t)=\left[ \begin{array}{ccccc}
I_{n\times n} & sI_{n\times n} & 0_{n\times 1} & 0_{n\times 1} & 0_{n\times 1} \\
0_{n\times n} & I_{n\times n} & 0_{n\times 1} & 0_{n\times 1} & 0_{n\times 1} \\
a(t+s,t) & d(t,s) & 1 & s & s^2/2 \\
0_{1\times n} & a(t+s,t) & 0 & 1 & s \\
0_{1\times n} & 0_{1\times n} & 0 & 0 & 1
\end{array} \right]
\end{equation}}
with $a(t+s,t):=x^{\top}(t+s)-x^{\top}(t)$, $d(t,s):=s\big(a(t+s,t)-\sum_{i=1}^l\alpha_i z_i^{\top}\big)$. We note that $|a(t+s,t)|$ is uniformly bounded and that $|d(t,s)|=O(s)$.\\
From \eqref{matrixM} and \eqref{matrixPhi}, we obtain
\vspace{-0.3cm}
{\small
\vspace{-0.5cm}
\begin{equation} \label{MPhix}
\begin{array}{l}
M(t+s)\phi(t+s,t)x=\\
\left[ \begin{array}{c}
a(t+s,t)x_1+d(t,s)x_2+x_3+sx_4+\frac{s^2}{2}x_5 \\
D(\alpha)Z^{\top}(x_1+sx_2) \\
\begin{array}{l}
u^{\top}(t+s)(x_1+sx_2)-(\sum_{i=1}^l\alpha_i z_i^{\top})x_2+a(t+s,t)x_2 \\+x_4+sx_5 \end{array} \\
D(\alpha)Z^{\top}x_2 \\
\dot{u}^{\top}(t+s)(x_1+sx_2)+2u^{\top}(t+s)x_2+x_5
\end{array} \right]
\end{array}
\end{equation}}
with $x_i$ ($i=1,\hdots,5$) denoting the $ith$ vector-part component of the vector $x\in \rr^{2n+3}$.\\
From now on, and for the sake of notation conciseness, we define $y(t,s,x):=M(t+s)\phi(t+s,t)x$. We make a proof by contradiction. Let us thus assume that the uniform observability property is not satisfied. Then, according to Proposition  \ref{proposition2} there exists a sequence $\{t_p\}_{p\in \nn}$ and $x \in \Sn^{2n+2}$ such that
$\lim_{p \rightarrow +\infty}\int_{0}^{\bar{\delta}+\delta}|y(t_p,s,x)|^2ds=0$ with $\bar{\delta}$ positive and as large as desired.\\
Since $\frac{s^2}{2}x_5$ dominates all other terms in the first component of $y(t,s,x)$ by a factor $s$, when $s$ is large, the satisfaction of the condition in Proposition \ref{proposition2} implies that $|x_5|$ has to become small as fast as $1/\bar{\delta}$ when $\bar{\delta}$ becomes large. We may thus assume from now on that $|x_5|\ll 1$.\\
From \eqref{MPhix}, $|y(t_p,s,x)|^2 \geq |DZ^{\top}(\alpha)x_2|^2$, so that, in view of the assumption \eqref{pecondition}, the satisfaction of the condition in Proposition \ref{proposition2} implies that $DZ^{\top}(\alpha)x_2=0$. Then, from \eqref{MPhix}, $|y(t_p,s,x)|^2 \geq |DZ^{\top}(\alpha)x_1|^2$, which in turn implies that $DZ^{\top}(\alpha)x_1=0$.\\
 Now, let us write the fifth component of $y(t,s,x)$ as $(\dot{u}^{\top}(t+s)x_2)s+r_1(t,s,x)$ with $r_1(t,s,x)\equiv \dot{u}^{\top}(t+s)x_1+2u^{\top}(t+s)x_2+x_5$. Then $|y(t,s,x)|^2\geq 0.5~(\dot{u}^{\top}(t+s)x_2)^2s^2-r_1(t,s,x)^2$. Therefore
\begin{equation} \label{eq1}
\begin{array}{l}
\int_{0}^{\bar{\delta}+\delta}|y(t,s,x)|^2ds \geq \\
\int_{\bar{\delta}}^{\bar{\delta}+\delta} \big( 0.5~(\dot{u}^{\top}(t+s)x_2)^2s^2-r_1(t,s,x)^2 \big)ds
\end{array}
\end{equation}
with
\[
\int_{\bar{\delta}}^{\bar{\delta}+\delta}(\dot{u}^{\top}(t+s)x_2)^2s^2ds \geq \bar{\delta}^2 \int_{\bar{\delta}}^{\bar{\delta}+\delta}(\dot{u}^{\top}(t+s)x_2)^2ds
\]
\[
\int_{\bar{\delta}}^{\bar{\delta}+\delta}|r_1(t,s,x)|^2ds \leq \delta R_1^2
\]
with $R_1$ denoting an upperbound of $|r_1(t,s,x)|$, using the fact that all terms involved in $r_1(t,s,x)$ and are uniformly bounded by assumption. Using these bounds in \eqref{eq1}
yields
\[
\int_{0}^{\bar{\delta}+\delta}|y(t,s,x)|^2ds \geq
0.5~\bar{\delta}^2 \int_{t+\bar{\delta}}^{t+\bar{\delta}+\delta}(\dot{u}^{\top}(s)x_2)^2ds-\delta R_1^2
\]
and, using the persistent excitation assumption \eqref{pecondition} according to which $\int_{t_p+\bar{\delta}}^{t_p+\bar{\delta}+\delta}(\dot{u}^{\top}(s)x_2)^2ds \geq \delta \mu |x_2|^2$ when $DZ^{\top}(\alpha)x_2=0$,
\[
\int_{0}^{\bar{\delta}+\delta}|y(t_p,s,x)|^2ds \geq
0.5~\bar{\delta}^2\delta \mu |x_2|^2-\delta R_1^2
\]
This latter relation shows that the convergence of $\int_{0}^{\bar{\delta}+\delta}|y(t_p,s,x)|^2ds$ to zero when $\bar{\delta}$ is large implies that $|x_2|$ must become small as fast than $1/\bar{\delta}$. We may thus also assume from now on that $|x_2|\ll 1$.\\
So far we have proven that by taking $\bar{\delta}$ very large, then the satisfaction of the condition in Proposition \ref{proposition2} implies that $|x_2|$ and $|x_5|$ are very small. We also know that $DZ^{\top}(\alpha)x_1=DZ^{\top}(\alpha)x_2=0$. Now, the convergence of $\int_{0}^{\bar{\delta}+\delta}|y(t_p,s,x)|^2ds$ to zero also implies the convergence of $\int_{0}^{\delta}|y(t_p,s,x)|^2ds$ to zero. Let us now rewrite the fifth component of $y(t_p,s,x)$ as $\dot{u}^{\top}(t_p+s)x_1+r_2(t_p,s,x)$ with $r_2(t_p,s,x)=\dot{u}^{\top}(t_p+s)sx_2+2u^{\top}(t_p+s)x_2+x_5$. Then
\[
\int_{0}^{\delta}|y(t_p,s,x)|^2ds \geq 0.5 \int_{0}^{\delta}|\dot{u}^{\top}(t_p+s)x_1|^2ds-\delta R_2^2
\]
with $R_2$ denoting an upperbound of $|r_2(t_p,s,x)|^2$ on the interval $s\in [0,\delta]$. Since $|x_2|$ and $|x_5|$ are as small as desired by choosing $\bar{\delta}$ as large as necessary, $R_2$ is itself as small as desired. And so is also $\delta R_2^2$. Using the persistent excitation assumption \eqref{pecondition} according to which
$\int_{t_p}^{t_p+\delta}(\dot{u}^{\top}(s)x_1)^2ds \geq \delta \mu |x_1|^2$ when $DZ^{\top}(\alpha)x_1=0$, we deduce that
\[
\int_{0}^{\delta}|y(t_p,s,x)|^2ds \geq 0.5 \delta \mu |x_1|^2-\delta R_2^2
\]
Therefore the convergence of $\int_{0}^{\delta}|y(t_p,s,x)|^2ds$ to zero implies that $|x_1| \ll 1$ when $\bar{\delta}$ is large.\\
Let us now consider the third component of $y(t_p,s,x)$, which is equal to $x_4+r_3(t_p,s,x)$ with $r_3(t_p,s,x)=u^{\top}(t+s)(x_1+sx_2)-(\sum_{i=1}^l\alpha_i z_i^{\top})x_2+a(t+s,t)x_2+sx_5$. Because $|x_1|$, $|x_2|$ and $|x_5|$ are as small as desired by choosing $\bar{\delta}$ as large as necessary, $|r_3(t_p,s,x)|$ is upperbounded on the interval $s\in [0,\delta]$ by a positive number $R_3$ which is itself as small as desired. Because $\int_{0}^{\delta}|y(t_p,s,x)|^2ds\geq 0.5x_4^2-\delta R_3^2$, the satisfaction  of the condition in Proposition \ref{proposition2} also implies $|x_4| \ll 1$ when $\bar{\delta}$ is large enough.\\
By considering the first component of $y(t_p,s,x)$, which is equal to $x_3+r_4(t_p,s,x)$ with $r_4(t_p,s,x)=a(t+s,t)x_1+d(t,s)x_2+sx_4+\frac{s^2}{2}x_5$ being very small on the interval $s\in [0,\delta]$ when $\bar{\delta}$ is large enough, we similarly show that the satisfaction of the condition in Proposition \ref{proposition2} implies $|x_3|\ll 1$ when $\bar{\delta}$ is large enough.\\
Because all the components of $x$ are small, the norm of $x$ must be smaller than one to satisfy the condition in  Proposition \ref{proposition2} when $\bar{\delta}$ is large enough. Because the satisfaction of this condition also requires the norm of $|x|$ being equal to one, we obtain a contradiction which finally proves that the assumption of non-uniform observability of the system does not hold.

\bibliographystyle{plain}        
\bibliography{bibfile3}
\end{document}